\pdfoutput=1

\RequirePackage{fix-cm}

\documentclass[twocolumn]{svjour3}

\smartqed

\usepackage{graphicx,amssymb}
\usepackage{epsf,psfig}
\usepackage{txfonts}
\usepackage{natbib}
\usepackage[unicode]{hyperref}

\hypersetup{pdftitle = The title of my PDF, pdfauthor = My name, pdfsubject= The subject, pdfkeywords = keyword1 keyword2 keyword3}
\hypersetup{colorlinks = true, linkcolor = blue, anchorcolor = red, citecolor = blue, filecolor = red, pagecolor = red, urlcolor = blue}

\journalname{Astrophysics and Space Science}

\begin{document}

\title{How does the oblateness coefficient influence the nature of orbits in the restricted three-body problem?}

\author{Euaggelos E. Zotos}

\institute{Department of Physics, School of Science, \\
Aristotle University of Thessaloniki, \\
GR-541 24, Thessaloniki, Greece \\
Corresponding author's email: {evzotos@physics.auth.gr}}

\date{Received: 30 May 2015 / Accepted: 2 July 2015 / Published online: 14 July 2015}

\titlerunning{Influence of the oblateness coefficient in the restricted three-body problem}

\authorrunning{Euaggelos E. Zotos}

\maketitle

\begin{abstract}
We numerically investigate the case of the planar circular restricted three-body problem where the more massive primary is an oblate spheroid. A thorough numerical analysis takes place in the configuration $(x,y)$ and the $(x,E)$ space in which we classify initial conditions of orbits into three categories: (i) bounded, (ii) escaping and (iii) collisional. Our results reveal that the oblateness coefficient has a huge impact on the character of orbits. Interpreting the collisional motion as leaking in the phase space we related our results to both chaotic scattering and the theory of leaking Hamiltonian systems. We successfully located the escape as well as the collisional basins and we managed to correlate them with the corresponding escape and collision times. We hope our contribution to be useful for a further understanding of the escape and collision properties of motion in this interesting version of the restricted three-body problem.


\end{abstract}

\section{Introduction}
\label{intro}

One of most interesting topics in nonlinear dynamics is the issue of leaking or escaping orbits (e.g., \citet{C90,CK92,CKK93,STN02,APT13,NKL07}). Over the years many studies have been devoted on Hamiltonian systems with escapes (e.g., \citet{BBS09,EP14,KSCD99,NH01,Z14a,Z14b,Z15a}). However the topic of escaping orbits in Hamiltonian systems is by far less explored than the closely related problem of chaotic scattering, where a test particle coming from infinity approaches and then scatters off a complex potential (e.g., \citet{BOG89,BTS96,BST98,JLS99,JMS95,JT91,SASL06,SSL07}).

Undoubtedly, one of the most important topics in dynamical astronomy is the classical restricted three-body problem (RTBP).
The RTBP describes the motion of a test body with an infinitesimal mass moving under the gravitational effects of a two primary bodies with finite masses which move in circular orbits around their common center of mass \citep{S67}. It is assumed that the third test body does not influence the motion of the two primary bodies. This problem has many applications in celestial mechanics, stellar systems, artificial satellites, galactic dynamics, chaos theory and molecular physics and therefore is still a very active and stimulating field of research.

Generally, the shapes of the two primary bodies in the classical version of the RTBP are assumed to be spherically symmetric. However, we found in nature that several celestial bodies, such as Saturn and Jupiter are sufficiently oblate \citep{BPC99} or even triaxial. Furthermore, irregular shapes are also possible, especially in the case of minor planets (e.g., Ceres) and meteoroids \citep{MWF87,NC08}. The oblateness or triaxiality of a celestial body can produce perturbation deviations from the two-body motion. The study of oblateness coefficient includes the series of works of \citet{BS12,MPP96,MRVK00,KMP05,KDP06,KPP08,KGK12,PPK12,SSR79,SSR86,SRS88,SRS97,S81,S87,S89,S90,SL12,SL13} by considering the more massive primary as an oblate spheroid with its equatorial plane co-incident with the plane of motion of the primaries.

All planets of the Solar System are in fact not spherically symmetric but oblate spheroidals. Consequently the exact shape of the planets and their oblateness must be taken into account when someone wants to model a space flight mission. In the work of \citet{OV03} it was proved that the theoretical data from the celestial systems Saturn-Tethys-satellite and Saturn-Dione-satellite are much more accurate with respect with the corresponding observational data when the oblateness of Saturn is taken into consideration. Similarly, the oblateness of the Neptune was included in the work of \citet{SYC08} regarding the dynamics of a spacecraft in the Neptune-Triton system.

In the present paper we continue the work initiated in \citet{Z15b} following the same numerical techniques. Our aim is to numerically investigate how the oblateness coefficient influences the nature of orbits in the restricted three-body problem. The paper is organized as follows: In Section \ref{mod} we describe the properties of the considered dynamical model along with some necessary theoretical details. All the computational methods we used in order to determine the character of orbits are explained in Section \ref{cometh}. In the following Section, we conduct a thorough numerical investigation revealing the overall orbital structure (bounded regions and basins of escape/collision) of the system and how it is affected by the value of the oblateness coefficient with respect to the total orbital energy. Our paper ends with Section \ref{conc}, where the main conclusions of this work are given.

\section{Details of the dynamical model}
\label{mod}

It would be very illuminating to briefly recall the basic properties and some aspects of the planar circular restricted three-body problem \citep{S67}. The two primaries move on circular orbits with the same Kepler frequency around their common center of gravity, which is assumed to be fixed at the origin of the coordinates. The third body (test particle with mass much smaller than the masses of the primaries) moves in the same plane under the gravitational field of the two primaries. The non-dimensional masses of the two primaries are $1-\mu$ and $\mu$, where $\mu = m_2/(m_1 + m_2)$, with $m_1 > m_2$ is the mass ratio. We consider an intermediate value of the mass ratio, that is $\mu = 1/10$. This value remains constant throughout the paper.

We choose as a reference frame a rotating coordinate system where the origin is at (0,0), while the centers $C_1$ and $C_2$ of the two primary bodies are located at $(-\mu, 0)$ and $(1-\mu,0)$, respectively. The total time-independent effective potential according to \citet{SSR76} is
\begin{equation}
V(x,y) = - \frac{\mu}{r_2} - \frac{(1 - \mu)}{r_1} - \frac{(1 - \mu)A_1}{2r_1^3} - \frac{n^2}{2}\left( x^2  + y^2 \right),
\label{pot}
\end{equation}
where
\[
r_1 = \sqrt{\left(x + \mu\right)^2 + y^2},
\]
\[
r_2 = \sqrt{\left(x + \mu - 1\right)^2 + y^2},
\]
\begin{equation}
n^2 = 1 + \frac{3 A_1}{2},
\label{dist}
\end{equation}
are the distances to the respective primaries and the angular velocity $(n)$, while $A_1$ is the oblateness coefficient which is defined as
\begin{equation}
A_1 = \frac{(RE)^2 - (RP)^2}{5R^2},
\label{obl}
\end{equation}
where $RE$ and $RP$ are the equatorial and polar radius, respectively of the oblate primary, while $R$ is the distance between the centers of the two primaries. We consider values of the oblateness coefficient is in the interval $[0,0.1]$ (see e.g., \citep{KDP06,PK06}), while we study the effect of oblateness up to the linear coefficient $J_2$ only \citep{A12}.

The human species stands on the edge of a new frontier, the transition from a planet-bound to a space-faring civilization. The expansion into the Solar System, in terms of dynamics of artificial satellites, requires the formulation of new models that include the effects of some of the perturbing forces on the satellite. Therefore, the problem is not only of mathematical interest but has astrophysical applications. The most striking example of perturbations arising from the oblateness in the Solar System is the orbit of the fifth satellite of Jupiter, Amalthea. This planet is so oblate and the satellite's orbit is so small that its line of apsides advances about $900^{\circ}$ in a year (e.g., \citet{M14}). It should be pointed out that in general terms the values of the oblateness in the Solar System are relatively low (see e.g., \citet{MW14} for updated values of the oblateness coefficient). There are however close-in extrasolar planetary systems with hot Jupiters in which the oblateness coefficient varies in the interval [0.001, 0.01]. In this work we consider even higher values of the oblateness coefficient $(A_1 < 0.1)$ because in the near future it is possible new highly oblate extrasolar planets to be discovered.

The scaled equations of motion describing the motion of the test body in the corotating frame read \citep{SSR76}
\[
\ddot{x} = 2n\dot{y} - \frac{\partial V(x,y)}{\partial x},
\]
\begin{equation}
\ddot{y} = - 2n\dot{x} - \frac{\partial V(x,y)}{\partial y}.
\label{eqmot}
\end{equation}
The dynamical system (\ref{eqmot}) admits the well know Jacobi integral
\begin{equation}
J(x,y,\dot{x},\dot{y}) = \frac{1}{2} \left(\dot{x}^2 + \dot{y}^2 \right) + V(x,y) = E,
\label{ham}
\end{equation}
where $\dot{x}$ and $\dot{y}$ are the velocities, while $E$ is the numerical value of the orbital energy which is conserved and defines a three-dimensional invariant manifold in the total four-dimensional phase space. Thus, an orbit with a given value of it's energy integral is restricted in its motion to regions in which $E \leq V(x,y)$, while all other regions are forbidden to the test body. It is widely believed that $J$ is the only independent integral of motion for the RTBP system \citep{P93}. The value of the energy $E$ is related with the Jacobi constant by $C = - 2E$.

\begin{figure*}[!tH]
\centering
\resizebox{\hsize}{!}{\includegraphics{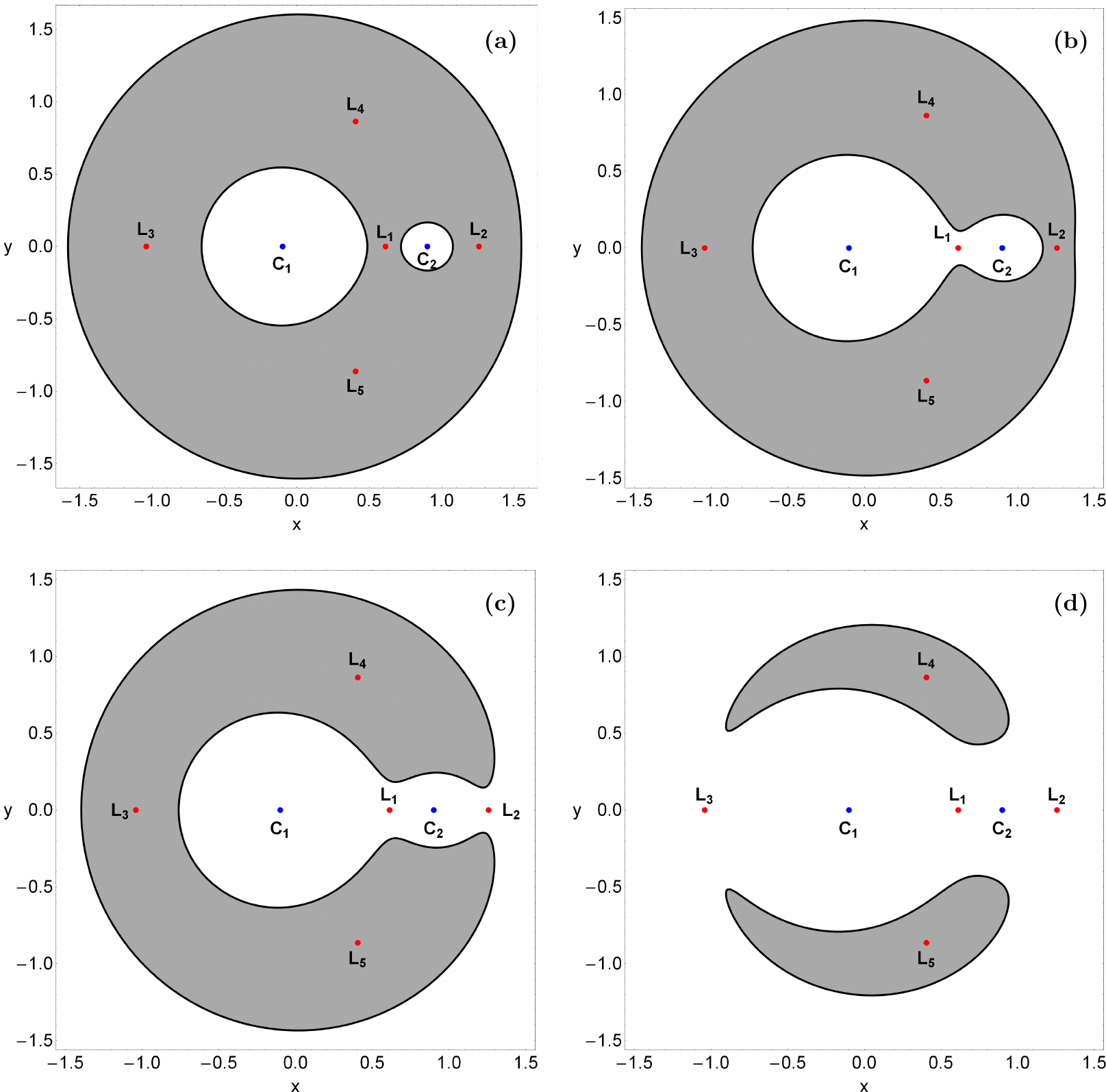}}
\caption{Four possible Hill's regions configurations for the RTBP system when $A_1 = 0.01$. The white domains correspond to the Hill's regions, gray shaded domains indicate the forbidden regions, while the thick black lines depict the Zero Velocity Curves (ZVCs). The red dots pinpoint the position of the Lagrangian points, while the positions of the centers of the two primary bodies are indicated by blue dots. (a-upper left): $E = -1.92$; (b-upper right): $E = -1.78$; (c-lower left): $E = -1.73$; (d-lower right): $E = -1.55$.}
\label{isos}
\end{figure*}

The dynamical system has five equilibria known as Lagrangian points \citep{S67} at which
\begin{equation}
\frac{\partial V(x,y)}{\partial x} = \frac{\partial V(x,y)}{\partial y} = 0.
\label{lps}
\end{equation}
Three of them, $L_1$, $L_2$, and $L_3$, are collinear points located in the $x$-axis. The central stationary point $L_1$ is a local minimum of the potential $V(x,y)$. The stationary points $L_2$ and $L_3$ are saddle points. Let $L_2$ located at $x > 0$, while $L_3$ be at $x < 0$. The points $L_4$ and $L_5$ on the other hand, are local maxima of the gravitational potential, enclosed by the banana-shaped isolines. The Lagrangian points are very important especially for astronautical applications. This can be seen in the Sun-Jupiter system where several thousand asteroids (collectively referred to as Trojan asteroids), are in orbits of equilibrium points. In practice these Lagrangian points have proven to be very useful indeed since a spacecraft can be made to execute a small orbit about one of these equilibrium points with a very small expenditure of energy \citep{SL14}.

The projection of the four-dimensional phase space onto the configuration (or position) space $(x,y)$ is called the Hill's regions and is divided into three domains: (i) the interior region for $x(L_3) \leq x \leq x(L_2)$; (ii) the exterior region for $x < x(L_3)$ and $x > x(L_2)$; (iii) the forbidden regions. The boundaries of these Hill's regions are called Zero Velocity Curves (ZVCs) because they are the locus in the configuration $(x,y)$ space where the kinetic energy vanishes.

\begin{table*}[!ht]
\begin{center}
   \caption{The position of the Lagrangian points and the critical values of the total orbital energy for four values of the oblateness coefficient.}
   \label{table1}
   \setlength{\tabcolsep}{10pt}
   \begin{tabular}{@{}lcccc}
      \hline
      $A_1$ & $L_1$ & $L_2$ & $L_3$ & $L_4$ \\
      \hline
      0.000 & (0.60903511, 0) & (1.25969983, 0) & (-1.04160890, 0) & (0.40000000, 0.86602540) \\
      0.001 & (0.60934671, 0) & (1.25945159, 0) & (-1.04165743, 0) & (0.40049937, 0.86573689) \\
      0.010 & (0.61207238, 0) & (1.25724426, 0) & (-1.04208268, 0) & (0.40493832, 0.86315542) \\
      0.100 & (0.63363978, 0) & (1.23751047, 0) & (-1.04544100, 0) & (0.44448280, 0.83877200) \\
      \hline
      $A_1$ & $E_1$ & $E_2$ & $E_3$ & $E_4$ \\
      \hline
      0.000 & -1.79847661 & -1.73334221 & -1.54978907 & -1.45500000 \\
      0.001 & -1.80001655 & -1.73471117 & -1.55114179 & -1.45613246 \\
      0.010 & -1.81381551 & -1.74701526 & -1.56331597 & -1.46632128 \\
      0.100 & -1.94701666 & -1.86855958 & -1.68503212 & -1.56790343 \\
      \hline
   \end{tabular}
\end{center}
\end{table*}

The values of the Jacobi integral at the five Lagrangian points $L_i$, $i = 1,...,5$ are critical energy levels and they are denoted as $E_i$ (Note that $E_4 = E_5$). The structure of the equipotential surfaces strongly depends on the value of the energy. In particular, there are five distinct cases
\begin{itemize}
  \item $E < E_1$: All necks are closed, so there are only bounded and collisional basins.
  \item $E_1 < E < E_2$: Only the neck around $L_1$ is open thus allowing orbits to move around both primaries.
  \item $E_2 < E < E_3$: The neck around $L_2$ is open, so orbits can enter the exterior region and escape form the system.
  \item $E_3 < E < E_4$: The necks around both $L_2$ and $L_3$ are open, so orbits can escape through two different escape channels.
  \item $E > E_4$: The banana-shaped forbidden regions disappear and therefore, motion over the entire configuration $(x,y)$ space is possible.
\end{itemize}
In Fig. \ref{isos}(a-d) we present for $A_1 = 0.01$ the structure of the first four possible Hill's region configurations. We observe in Fig. \ref{isos}d the two openings (exit channels) at the Lagrangian points $L_2$ and $L_3$ through which the test body can enter the exterior region and then leak out. In fact, we may say that these two exits act as hoses connecting the interior region of the system where $x(L_3) \leq x \leq x(L_2)$ with the ``outside world" of the exterior region. The position of the Lagrangian points as well as the critical values of the energy are functions of the oblateness coefficient $A_1$ (e.g., \citet{SL12}). In Table \ref{table1} we provide the location of the Lagrangian points and the critical values of the total orbital energy when $A_1 = \{0, 0.001, 0.01, 0.1\}$.

\section{Computational methods and criteria}
\label{cometh}

The motion of the test third body is restricted to a three-dimensional surface $E = const$, due to the existence of the Jacobi integral. With polar coordinates $(r,\phi)$ in the center of the mass system of the corotating frame the condition $\dot{r} = 0$ defines a two-dimensional surface of section, with two disjoint parts $\dot{\phi} < 0$ and $\dot{\phi} > 0$. Each of these two parts has a unique projection onto the configuration $(x,y)$ space. In order to explore the orbital structure of the system we need to define samples of initial conditions of orbits whose properties will be identified. Fore this purpose, we define for several values of the total orbital energy $E$, as well as for the oblateness coefficient $A_1$ dense uniform grids of $1024 \times 1024$ initial conditions regularly distributed on the configuration $(x,y)$ space inside the area allowed by the value of the total orbital energy. Following a typical approach, the orbits are launched with initial conditions inside a certain region, called scattering region, which in our case is a square grid with $-2\leq x,y \leq 2$.

\begin{figure}[!tH]
\centering
\includegraphics[width=\hsize]{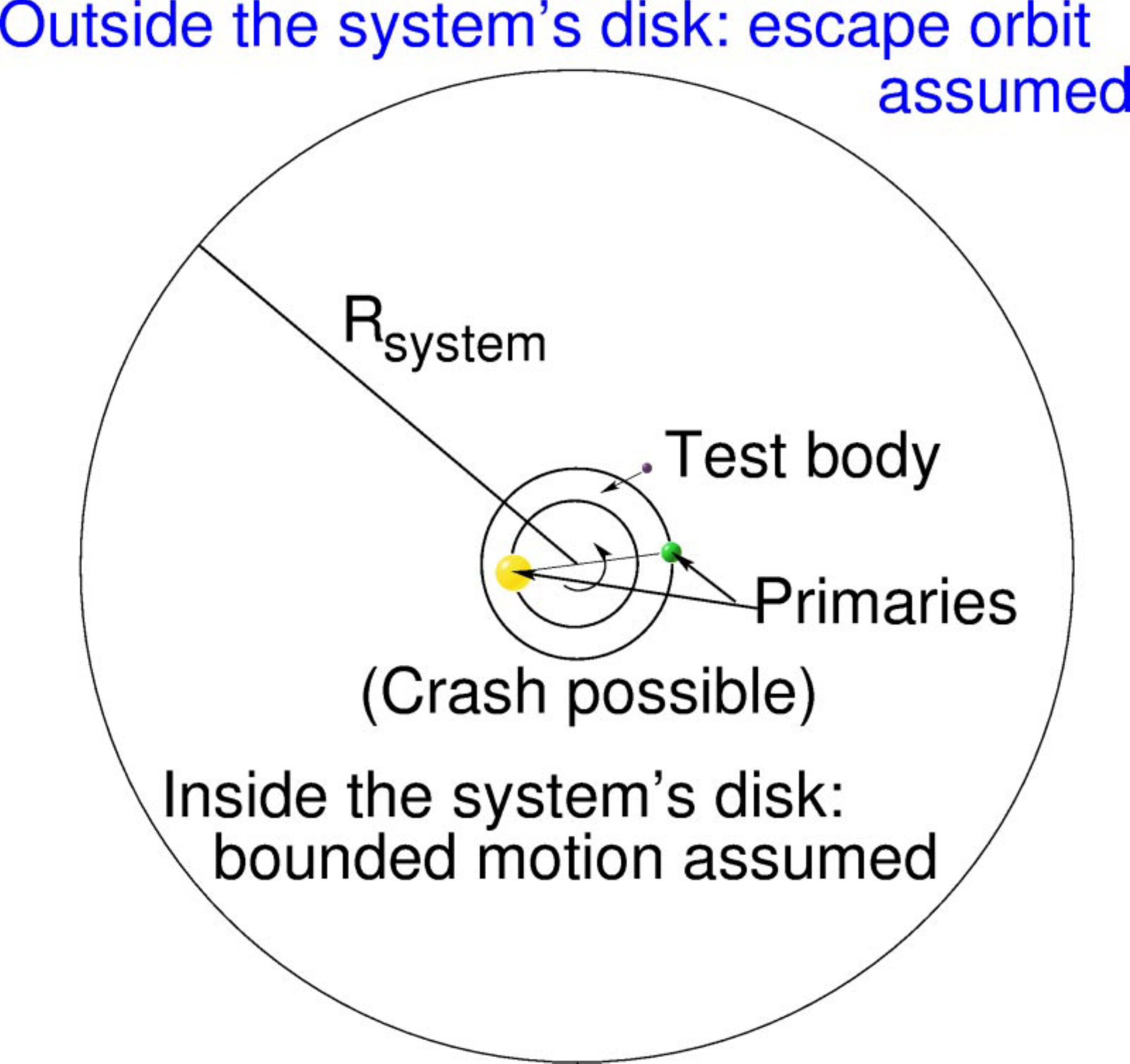}
\caption{Schematic picture of the three different types of motion. The motion is considered to be bounded if the test body stays confined for integration time $t_{\rm max}$ inside the system's disk with radius $R_d = 10$, while the motion is unbounded and the numerical integration stops when the test body crosses the system's disk with velocity pointing outwards. Collision or (crash) with one of the primaries occurs when the test body crosses the disk of radii $R_{m1}$ and $R_{m2}$ of the primaries.}
\label{crit}
\end{figure}

In the RTBP system the configuration space extends to infinity thus making the identification of the type of motion of the test body for specific initial conditions a rather demanding task. There are three possible types of motion for the test body: (i) bounded motion around one of the primaries, or even around both; (ii) escape to infinity; (iii) collision into one of the two primaries. Now we need to define appropriate numerical criteria for distinguishing between these three types of motion. The motion is considered as bounded if the test body stays confined for integration time $t_{\rm max}$ inside the system's disk with radius $R_d$ and center coinciding with the center of mass origin at $(0,0)$. Obviously, the higher the values of $t_{\rm max}$ and $R_d$ the more plausible becomes the definition of bounded motion and in the limit $t_{\rm max} \rightarrow \infty$ the definition is the precise description of bounded motion in a finite disk of radius $R_d$. Consequently, the higher these two values, the longer the numerical integration of initial conditions of orbits lasts. In our calculations we choose $t_{\rm max} = 10^4$ and $R_d = 10$ (see Fig. \ref{crit}) as in \citet{N04,N05} and \citet{Z15b}. We decided to include a relatively high disk radius $(R_d = 10)$ in order to be sure that the orbits will certainly escape from the system and not return back to the interior region. Furthermore, it should be emphasized that for low values of $t_{\rm max}$ the fractal boundaries of stability islands corresponding to bounded motion become more smooth. Moreover, an orbit is identified as escaping and the numerical integration stops if the test body body intersects the system's disk with velocity pointing outwards at a time $t_{\rm esc} < t_{\rm max}$. Finally, a collision with one of the primaries occurs if the test body, assuming it is a point mass, crosses the disk with radius $R_m$ around the primary. For the larger oblate primary we choose $R_{m_1} = 10^{-4}$. Generally, it is assumed that the radius of a celestial body (e.g., a planet) is directly proportional to the cubic root of its mass. Therefore, for the sake of simplicity of the numerical calculations we adopt the simple relation between the radii of the primaries
\begin{equation}
R_{m_2} = R_{m_1} \times \left(2\mu\right)^{1/3},
\label{radii}
\end{equation}
which was introduced in \citet{N05}. In \citet{N04,N05} it was shown that the radii of the primaries influence the area of collision and escape basins.

As it was stated earlier, in our computations, we set $10^4$ time units as a maximum time of numerical integration. The vast majority of escaping orbits (regular and chaotic) however, need considerable less time to escape from the system (obviously, the numerical integration is effectively ended when an orbit moves outside the system's disk and escapes). Nevertheless, we decided to use such a vast integration time just to be sure that all orbits have enough time in order to escape. Remember, that there are the so called ``sticky orbits" which behave as regular ones during long periods of time. Here we should clarify, that orbits which do not escape after a numerical integration of $10^4$ time units are considered as non-escaping or trapped.

The equations of motion (\ref{eqmot}) for the initial conditions of all orbits are forwarded integrated using a double precision Bulirsch-Stoer \verb!FORTRAN 77! algorithm (e.g., \citet{PTVF92}) with a small time step of order of $10^{-2}$, which is sufficient enough for the desired accuracy of our computations. Here we should emphasize, that our previous numerical experience suggests that the Bulirsch-Stoer integrator is both faster and more accurate than a double precision Runge-Kutta-Fehlberg algorithm of order 7 with Cash-Karp coefficients. Throughout all our computations, the Jacobian energy integral (Eq. (\ref{ham})) was conserved better than one part in $10^{-11}$, although for most orbits it was better than one part in $10^{-12}$. For collisional orbits where the test body moves inside a region of radius $10^{-2}$ around one of the primaries the Lemaitre's global regularization method is applied.

\section{Numerical results \& Orbit classification}
\label{numres}

The main numerical task is to classify initial conditions of orbits in the $\dot{\phi} < 0$ part\footnote{We choose the $\dot{\phi} < 0$ instead of the $\dot{\phi} > 0$ part simply because in \citet{Z15b} we seen that it contains more interesting orbital content.} of the surface of section $\dot{r} = 0$ into three categories: (i) bounded orbits; (ii) escaping orbits and (iii) collisional orbits. Moreover, two additional properties of the orbits will be examined: (i) the time-scale of collision and (ii) the time-scale of the escapes (we shall also use the terms escape period or escape rates). In this work we shall explore these dynamical quantities for various values of the total orbital energy, as well as for the oblateness coefficient $A_1$. In particular, we choose three energy levels which correspond to the last three Hill's regions configurations. The first two Hill's regions configurations contain only bounded and collisional orbits, so the orbital content is not so interesting. In the following color-coded grids (or orbit type diagrams - OTDs) each pixel is assigned a color according to the orbit type. Thus the initial conditions of orbits are classified into bounded orbits, unbounded or escaping orbits and collisional orbits. In this special type of Poincar\'{e} surface of section the phase space emerges as a close and compact mix of escape basins, collisional basins and stability islands. Our numerical calculations indicate that apart from the escaping and collisional orbits there is also a considerable amount of non-escaping orbits. In general terms, the majority of non-escaping regions corresponds to initial conditions of regular orbits, where an adelphic integral of motion is present, restricting their accessible phase space and therefore hinders their escape.

\subsection{Results for Region III: $E_2 < E < E_3$}

\begin{figure*}[!tH]
\centering
\resizebox{\hsize}{!}{\includegraphics{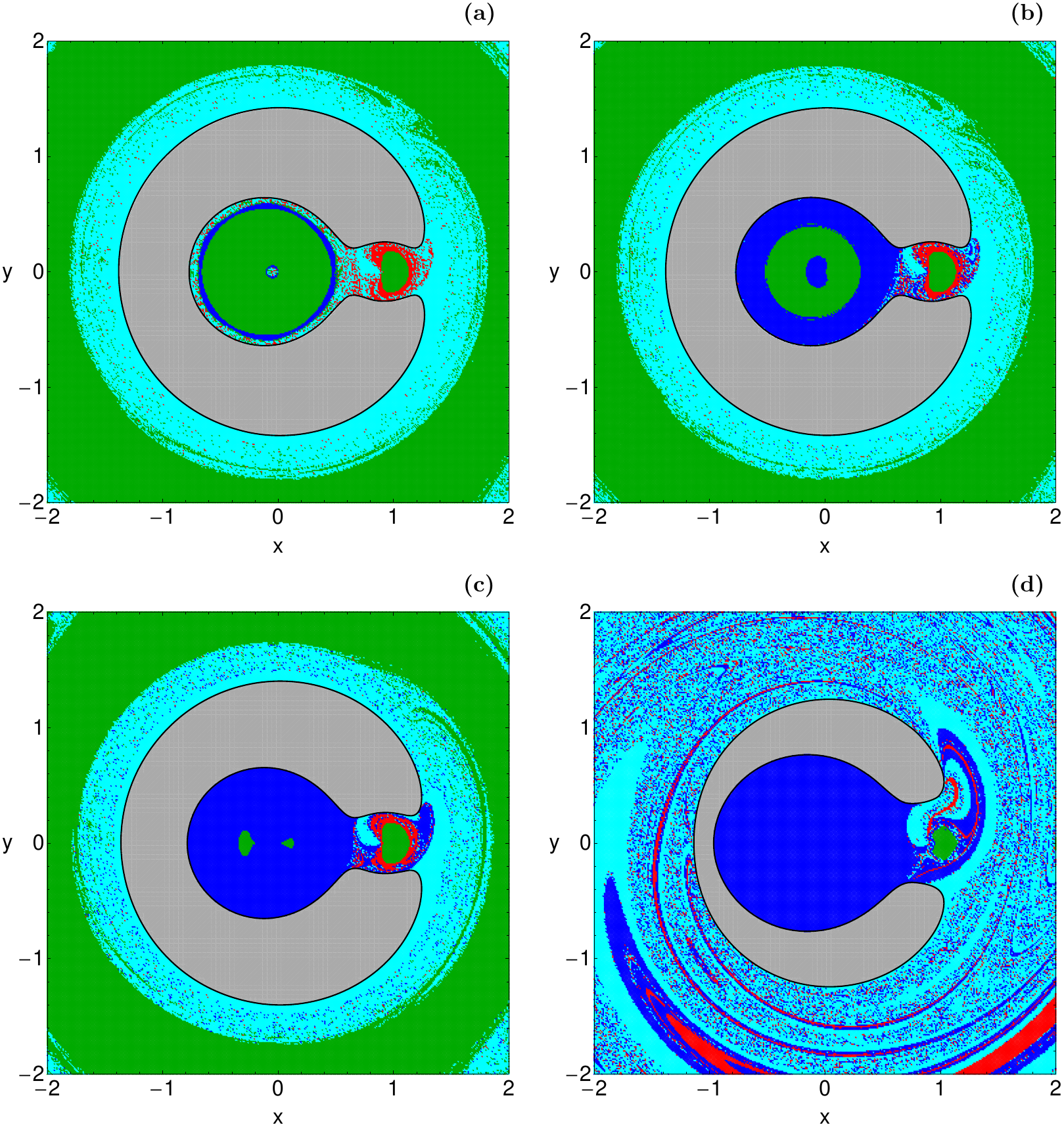}}
\caption{The orbital structure of the $\dot{\phi} < 0$ part of the surface of section $\dot{r} = 0$ when $E = -1.7$. (a-upper left): $A_1 = 0$; (b-upper right): $A_1 = 0.001$; (c-lower left): $A_1 = 0.01$; (d-lower right): $A_1 = 0.1$. The color code is the following: bounded orbits (green), collisional orbits to oblate primary 1 (blue), collisional orbits to primary 2 (red), and  escaping orbits (cyan).}
\label{E1}
\end{figure*}

\begin{figure*}[!tH]
\centering
\resizebox{\hsize}{!}{\includegraphics{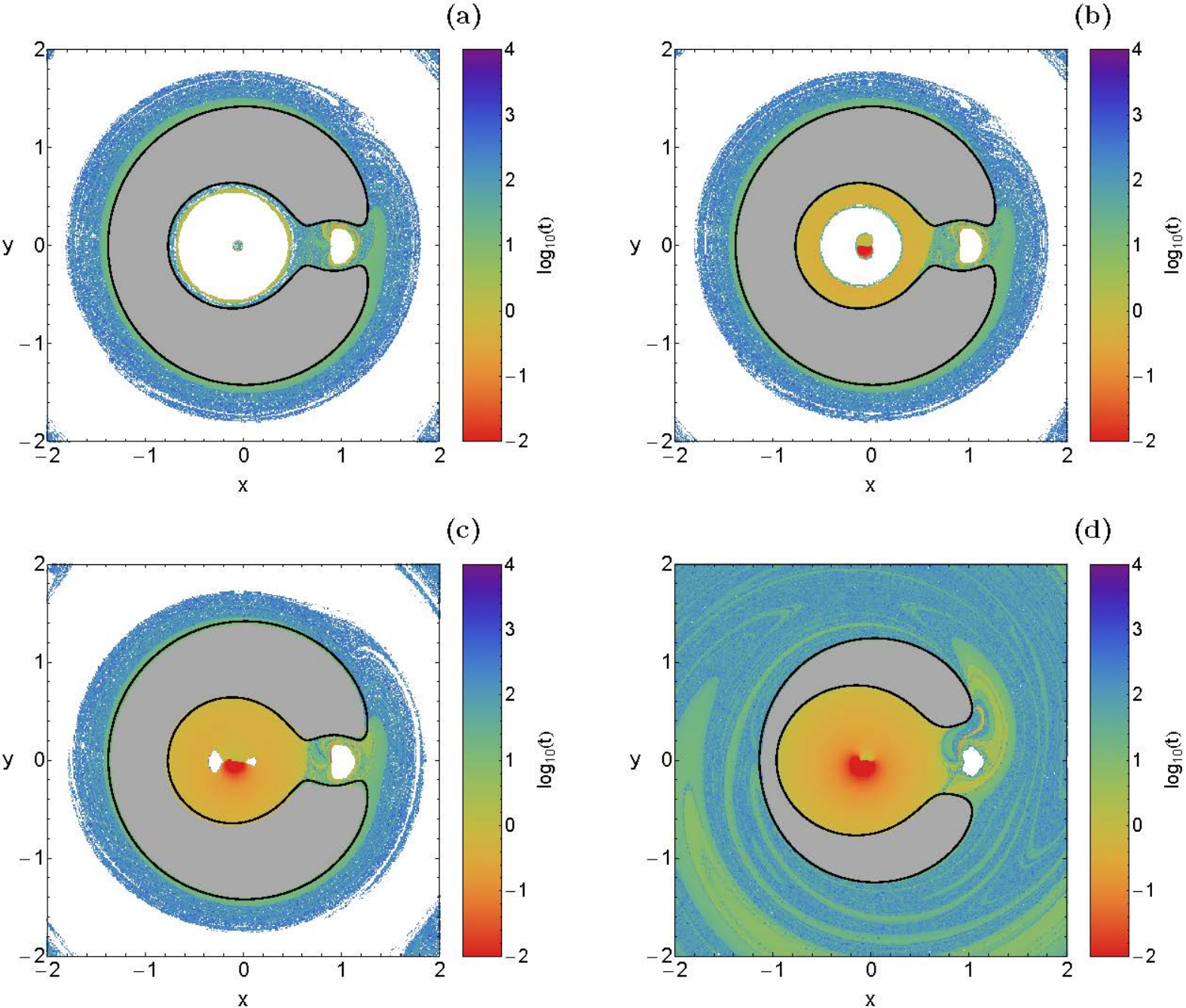}}
\caption{Distribution of the escape and collisional time of the orbits on the $\dot{\phi} < 0$ part of the surface of section $\dot{r} = 0$ when $E = -1.70$ for the values of the oblateness coefficient of Fig. \ref{E1}. The darker the color, the larger the escape/collisional time. Initial conditions of bounded regular orbits are shown in white.}
\label{E1t}
\end{figure*}

Our numerical exploration begins in the energy region III $(E_2 < E < E_3)$ and we choose the energy level $E = -1.70$. In Fig. \ref{E1} the OTD decompositions of the $\dot{\phi} < 0$ part of the surface of section $\dot{r} = 0$ reveal the orbital structure of the configuration $(x,y)$ space for four values of the oblateness coefficient. The black solid lines denote the ZVC, while the inaccessible forbidden regions are marked in gray. The color of a point represents the orbit type of a test body which has been launched with pericenter position at $(x,y)$. When the oblateness coefficient is zero $(A_1) = 0$ we observe in Fig. \ref{E1}a the followings: (i) around the centers of the two primaries there are stability islands, while in the exterior region another ring-shaped stability island is present. The stability islands located in the interior region correspond to regular orbits around one of the primary bodies, while the stability island in the exterior region is formed by initial conditions of regular orbits that circulate around both primaries; (ii) The region in the boundaries of the stability islands of the interior region is mainly occupied by orbits which lead to collision; (iii) Near the center of primary 1 we can identify a small hole which contains a mixture of collisional and escaping orbits; (iv) outside the forbidden regions there is a well-formed circular basin of escaping orbits. It is evident in Fig. \ref{E1}b that even a very low values of the oblateness coefficient $(A_1 = 0.001)$ is able to influence the orbital content. In particular, there are two main differences with respect to the case of zero oblateness shown in Fig. \ref{E1}a. First the area of the stability region around the center of the oblate primary is smaller, while the central hole contains now only initial conditions of collisional orbits. Moreover, the collisional basin around primary 1 is larger in size. The orbital structure changes even further when $A_1 = 0.01$ as we can see in Fig. \ref{E1}c. Almost all the area around the oblate primary 1 is occupied by initial conditions of collisional orbits, while the corresponding stability island splits into two small parts. It seems that so far the region around smaller primary 2 as well as the exterior region remain almost unperturbed by the shift on the value of the oblateness coefficient. Things however change drastically when the larger primary is highly oblate with $A_1 = 0.1$. It is seen in Fig. \ref{E1}d that the entire interior region around primary 1 is dominated by a large collisional basin. The exterior region on the other hand, is highly fractal, while the stability island of regular orbits circulating around both primaries is now absent. At least at this energy level our computations suggest that the increase in the value of the oblateness coefficient does not affect the stability island around primary 2, however at the highest studied value the collisional basin around this stability island is weakened. It should be noted that when $A_1 = 0.1$ several spiral collisional basins are inside the fractal exterior region.

The following Fig. \ref{E1t} shows how the escape and collisional times of orbits are distributed on the configuration $(x,y)$ space for the four values of the oblateness coefficient discussed in Fig. \ref{E1}. Light reddish colors correspond to fast escaping/collional orbits, dark blue/purple colors indicate large escape/collional times, while white color denote stability islands of regular motion. Note that the scale on the color bar is logarithmic. Inspecting the spatial distribution of various different ranges of escape time, we are able to associate medium escape time with the stable manifold of a non-attracting chaotic invariant set, which is spread out throughout this region of the chaotic sea, while the largest escape time values on the other hand, are linked with sticky motion around the stability islands of the two primary bodies. As for the collisional time we see that orbits with initial conditions very close to the vicinity of the center of the oblate primary 1 collide with it almost immediately, within the first time step of the numerical integration, while this phenomenon is not observed for the case of primary body 2 which is not oblate (see also \citet{Z15b}). Looking more carefully Fig. \ref{E1t}d we observe that when $A_1 = 0.1$ the area of the stability region around primary 2 (the only stability region that survives) is smaller with respect to the tree previous cases shown in Figs. \ref{E1t}(a-c). Thus we may say that high enough values of the oblateness coefficient influence also the stability region corresponding to smaller spherically symmetric primary.

\subsection{Results for Region IV: $E_3 < E < E_4$}

\begin{figure*}[!tH]
\centering
\resizebox{\hsize}{!}{\includegraphics{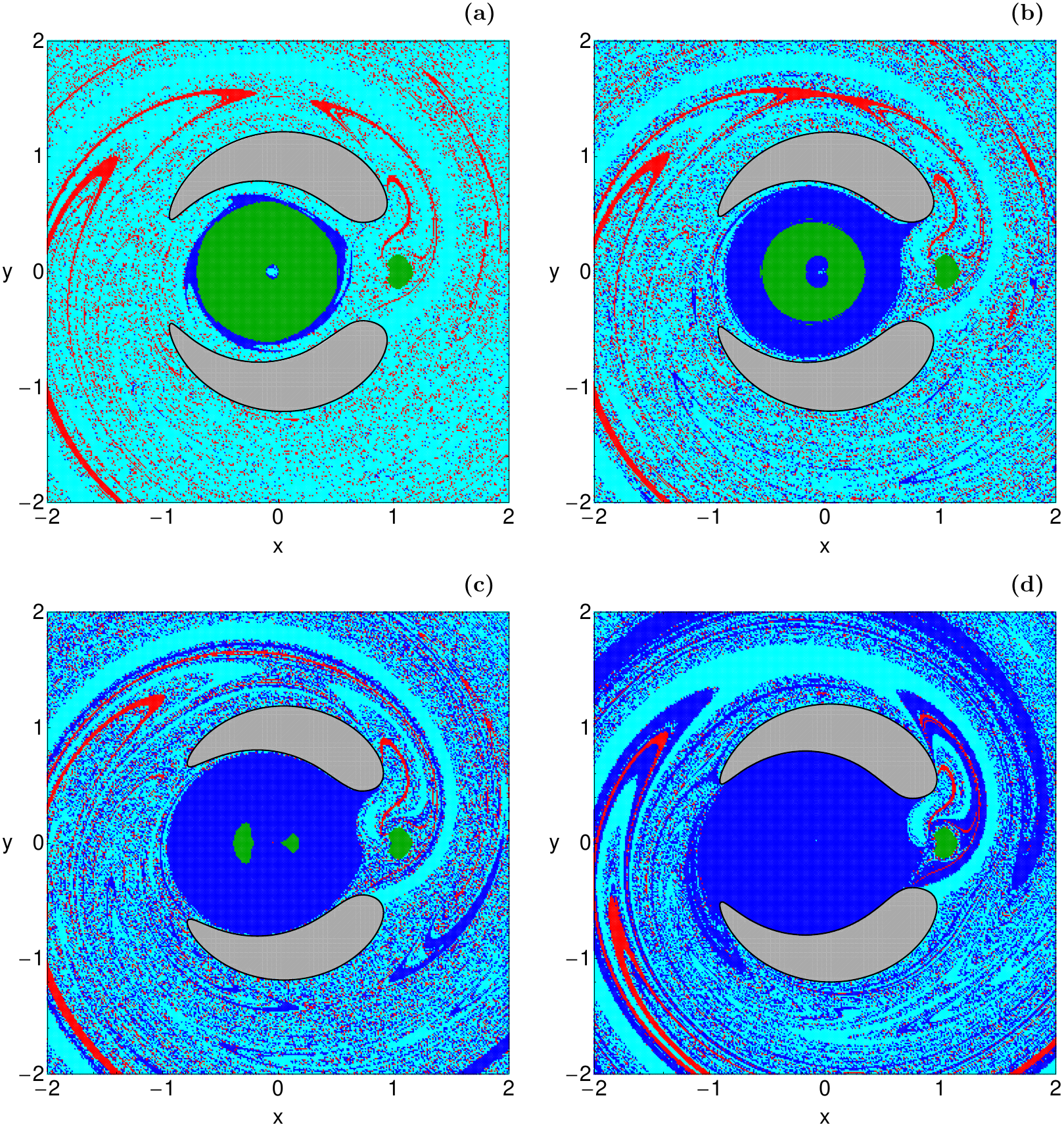}}
\caption{The orbital structure of the $\dot{\phi} < 0$ part of the surface of section $\dot{r} = 0$ when (a-c): $E = -1.54$ and (d-lower right): $E_1 = -1.67$. (a-upper left): $A_1 = 0$; (b-upper right): $A_1 = 0.001$; (c-lower left): $A_1 = 0.01$; (d-lower right): $A_1 = 0.1$. The color code is the same as in Fig. \ref{E1}.}
\label{E2}
\end{figure*}

\begin{figure*}[!tH]
\centering
\resizebox{\hsize}{!}{\includegraphics{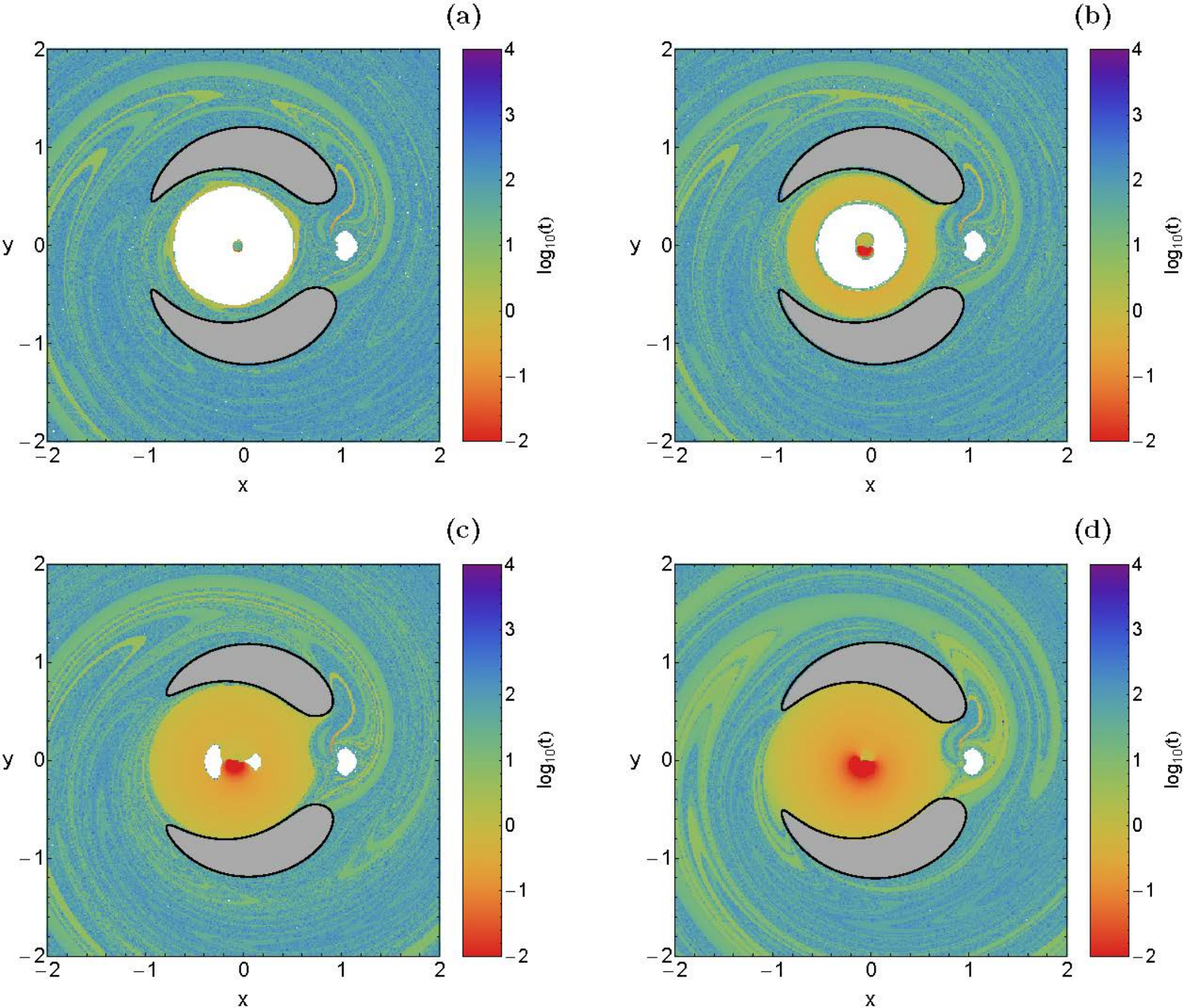}}
\caption{Distribution of the escape and collisional time of the orbits on the $\dot{\phi} < 0$ part of the surface of section $\dot{r} = 0$ when (a-c): $E = -1.54$ and (d-lower right): $E_1 = -1.67$ for the values of the oblateness coefficient of Fig. \ref{E2}.}
\label{E2t}
\end{figure*}

We continue our investigation in the energy region IV $(E_3 < E < E_4)$. It is evident from the critical energy values given in Table \ref{table1} that there is not a single energy level which we can use for all four cases regarding the value of the oblateness coefficient. Therefore for the first three cases, that is $A_1 = \{0, 0.001, 0.01\}$, we choose the value $E_1 = -1.54$, while for the last case $(A_1 = 0.1)$ we use the energy level $E_1 = -1.67$. The orbital structure of the configuration $(x,y)$ space is unveiled in Fig. \ref{E2} through the OTD decompositions of the $\dot{\phi} < 0$ part of the surface of section $\dot{r} = 0$. We observe in Fig. \ref{E2}a where $A_1 = 0$ that in this case the stability island in the exterior region is no longer present. The vast majority of the exterior region is occupied by initial conditions of escaping orbits, while inside the extended escape basin we identify delocalized initial conditions of orbits that collide to the oblate primary. On the other hand, the initial conditions of orbits which collide to primary 2 form well-defined spiral basins. The interior region is almost the same with that discussed earlier in Fig. \ref{E1}a, however there is a minor difference; the collisional basin around the stability island of primary 2 is absent now. With the introduction of the oblateness coefficient it is seen in Figs. \ref{E2}(b-d) that the interior region displays exactly the same behaviour as in the previous case. In the exterior region two phenomena take place as the value of the oblateness coefficient increases: (i) the amount of the escaping orbits decreases while at the same time the portion of the collisional orbits to oblate primary 1 increases and (ii) the fractality of the exterior region gradually decreases with increasing $A_1$ and several escape and collisional basins emerge.

The distribution of the escape and collisional times of orbits on the configuration space is shown in Fig. \ref{E2t}. One may observe that the results are very similar to those presented earlier in Fig. \ref{E1t}, where we found that orbits with initial conditions inside the escape and collisional basins have the smallest escape/collision rates, while on the other hand, the longest escape/collisional rates correspond to orbits with initial conditions in the fractal regions of the OTDs. Our calculations reveal, and this can be seen better in Figs. \ref{E2t}(a-d), that in this energy region the oblateness coefficient does not affect the size of the stability island of motion around primary 2.

\subsection{Results for Region V: $E > E_4$}

\begin{figure*}[!tH]
\centering
\resizebox{\hsize}{!}{\includegraphics{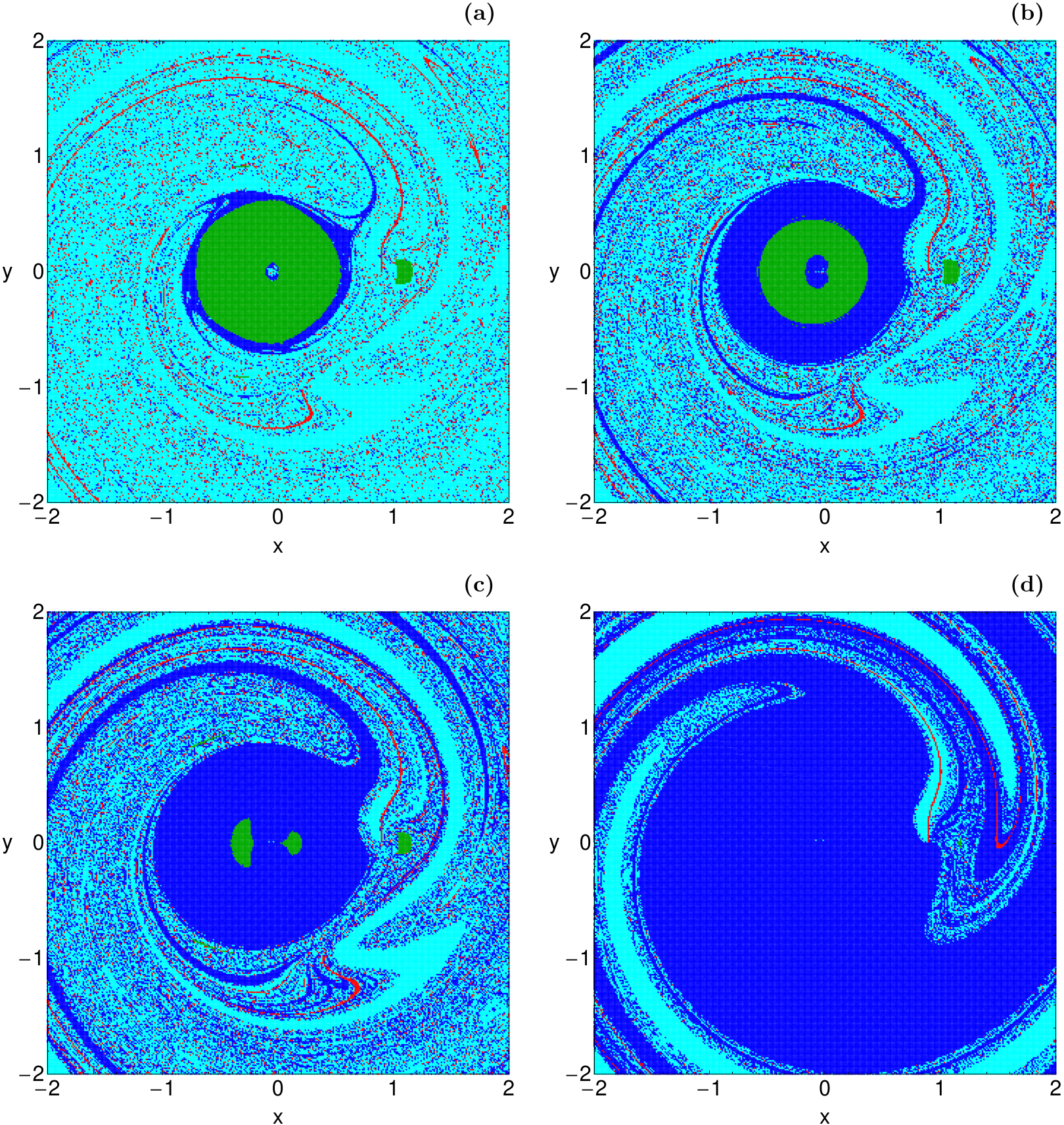}}
\caption{The orbital structure of the $\dot{\phi} < 0$ part of the surface of section $\dot{r} = 0$ when $E_1 = -1.45$. (a-upper left): $A_1 = 0$; (b-upper right): $A_1 = 0.001$; (c-lower left): $A_1 = 0.01$; (d-lower right): $A_1 = 0.1$. The color code is the same as in Fig. \ref{E1}.}
\label{E3}
\end{figure*}

\begin{figure*}[!tH]
\centering
\resizebox{\hsize}{!}{\includegraphics{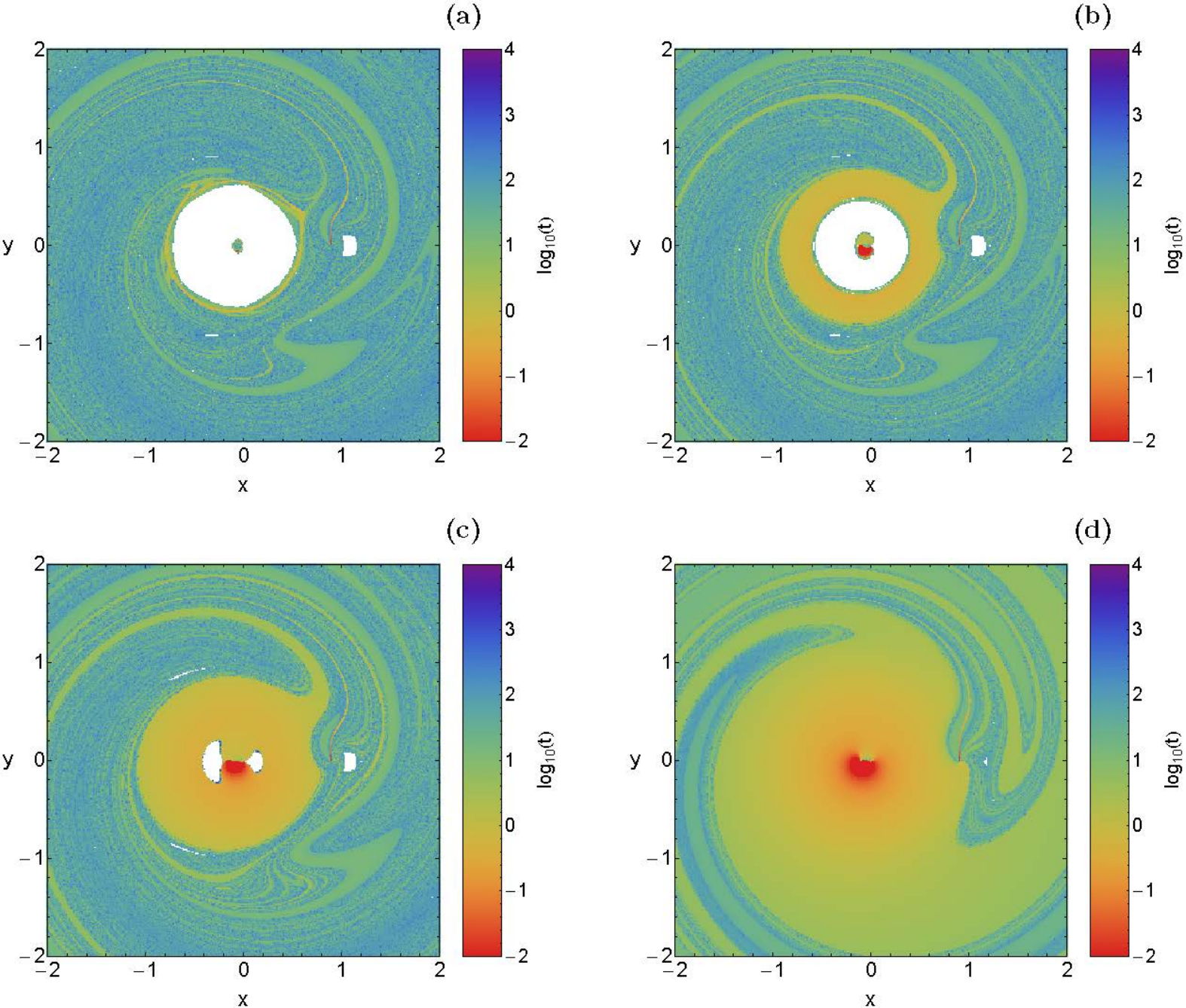}}
\caption{Distribution of the escape and collisional time of the orbits on the $\dot{\phi} < 0$ part of the surface of section $\dot{r} = 0$ when $E_1 = -1.45$ for the values of the oblateness coefficient of Fig. \ref{E3}.}
\label{E3t}
\end{figure*}

The last case under consideration involves the results in the energy region V where the entire configuration space is available for motion since the forbidden regions completely disappear. Once more, all the different aspects of the numerical approach remain exactly the same as in the two previously studied cases. Fig. \ref{E3} presents the orbital structure of the configuration space through the OTD decompositions of the $\dot{\phi} < 0$ part of the surface of section $\dot{r} = 0$. When both primaries are spherically symmetric, that is when $A_1 = 0$, we see in Fig. \ref{E3}a that most of the configuration $(x,y)$ plane is highly fractal, while a basin of collisional orbits is located around the stability island of primary 1. Around primary 2 there is another stability island but now the corresponding collisional basin identified in energy region III is not present now. In addition, the central hole inside the stability island of primary 1 still survives. At the lowest value of the oblateness coefficient $(A_1 = 0.001)$ the observed changes in Fig. \ref{E3}b are the same with those reported earlier in Fig. \ref{E2}b. The same applies in the case for $A_1 = 0.01$ where the results are shown in Fig. \ref{E3}c. Finally in Fig. \ref{E3}d where we have the scenario with the highest possible value of the oblateness coefficient $(A_1 = 0.1)$ it is seen that the vast majority of the $(x,y)$ plane is occupied by well-formed escape basins and collisional basins, while the fractal areas are confined to the boundaries between the several basins. It should be pointed out that in this case the area of the stability region around primary 2 is considerably reduced. Thus one may reasonably conclude that in the energy region V and for high enough values of $A_1$ the stability region around spherically primary 2 is highly affected by the oblateness coefficient.

In Fig. \ref{E3t} we depict the distribution of the escape and collisional times of orbits on the configuration space. One can see similar outcomes with that presented in the two previous subsections. At this point, we would like to emphasize that the basins of escape can be easily distinguished in Fig. \ref{E3t}, being the regions with intermediate colors indicating fast escaping orbits. Indeed, our numerical computations suggest that orbits with initial conditions inside these basins need no more than 10 time units in order to escape from the system. Furthermore, the collisional basins are shown with reddish colors where the corresponding collisional time is less than one time unit of numerical integration.

The OTDs shown in Figs. \ref{E1}, \ref{E2} and \ref{E3} have both fractal and non-fractal (smooth) boundary regions which separate the escape basins from the collisional basins. Such fractal basin boundaries is a common phenomenon in leaking Hamiltonian systems (e.g., \citet{BGOB88,dML99,dMG02,STN02,ST03,TSPT04}). In the RTBP system the leakages are defined by both escape and collision conditions thus resulting in three exit modes. However, due to the high complexity of the basin boundaries, it is very difficult, or even impossible, to predict in these regions whether the test body (e.g., a satellite, asteroid, planet etc) collides with one of the primary bodies or escapes from the dynamical system.

\subsection{An overview analysis}

\begin{figure*}[!tH]
\centering
\resizebox{\hsize}{!}{\includegraphics{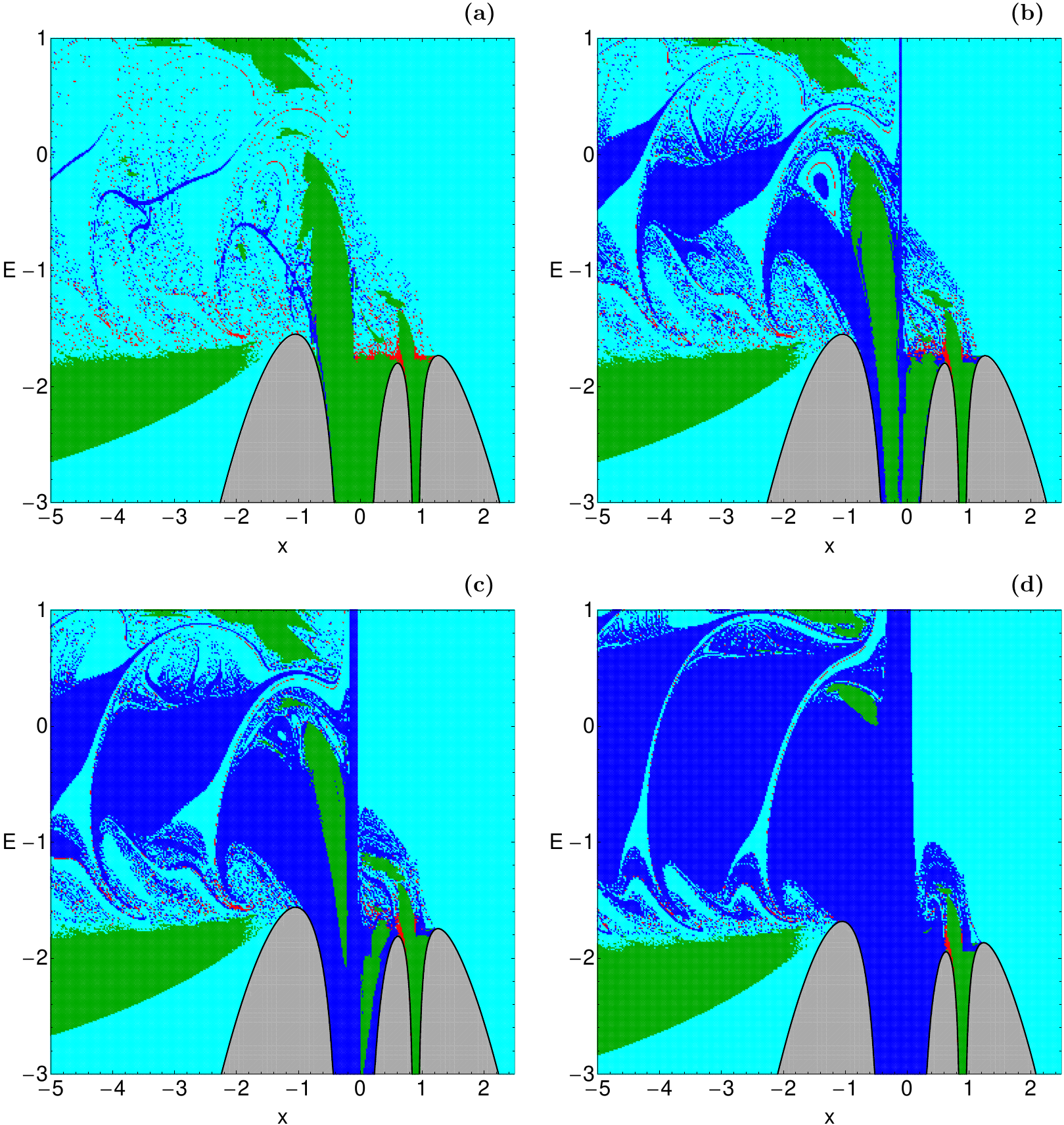}}
\caption{Orbital structure of the $(x,E)$ plane when (a-upper left): $A_1 = 0$; (b-upper right): $A_1 = 0.001$; (c-lower left): $A_1 = 0.01$; (d-lower right): $A_1 = 0.1$. The color code is the same as in Fig. \ref{E1}.}
\label{xE}
\end{figure*}

\begin{figure*}[!tH]
\centering
\resizebox{\hsize}{!}{\includegraphics{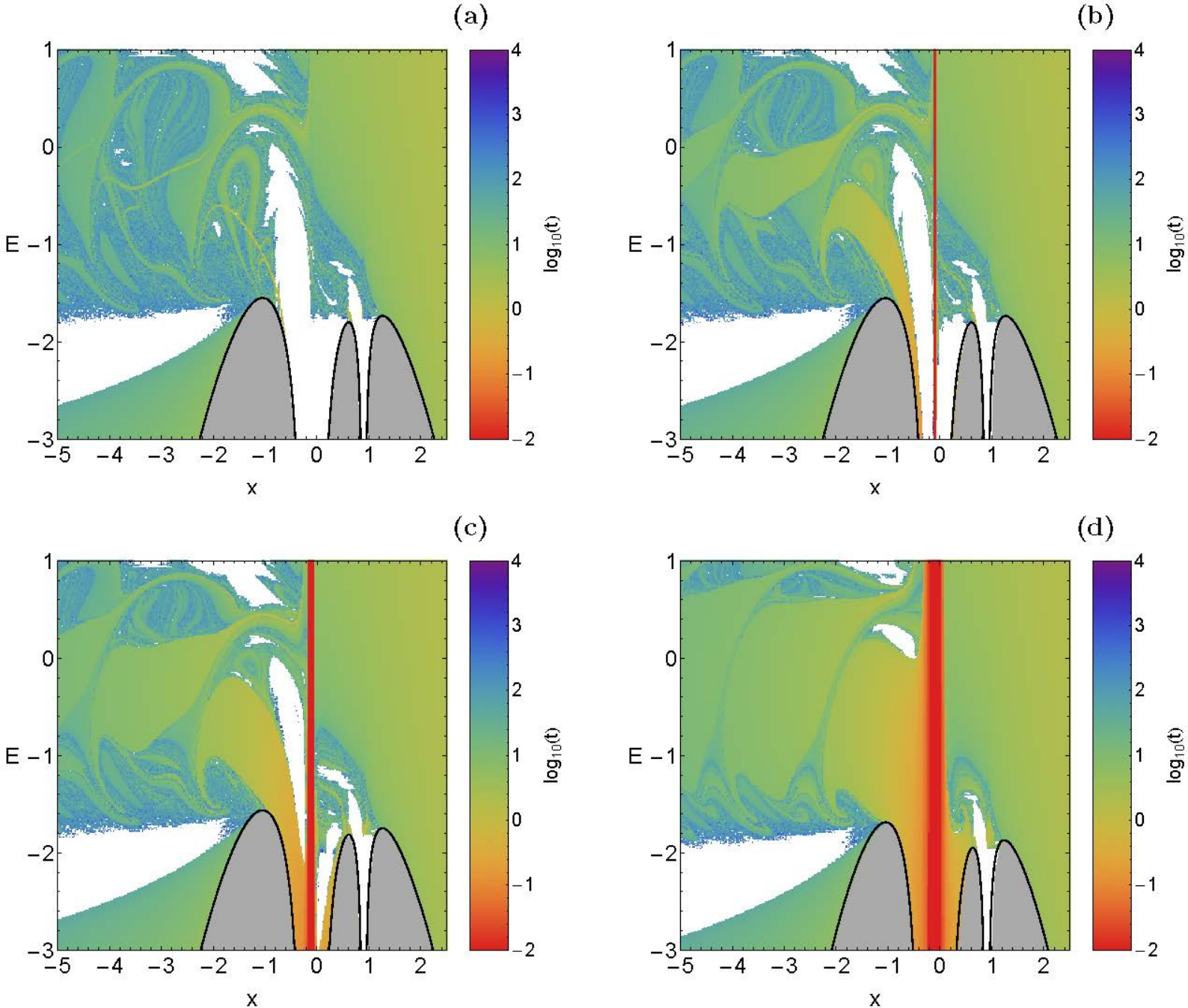}}
\caption{Distribution of the escape and collisional time of the orbits on the $(x,E)$ plane for the values of the oblateness coefficient of Fig. \ref{xE}.}
\label{xEt}
\end{figure*}

\begin{figure*}[!tH]
\centering
\resizebox{\hsize}{!}{\includegraphics{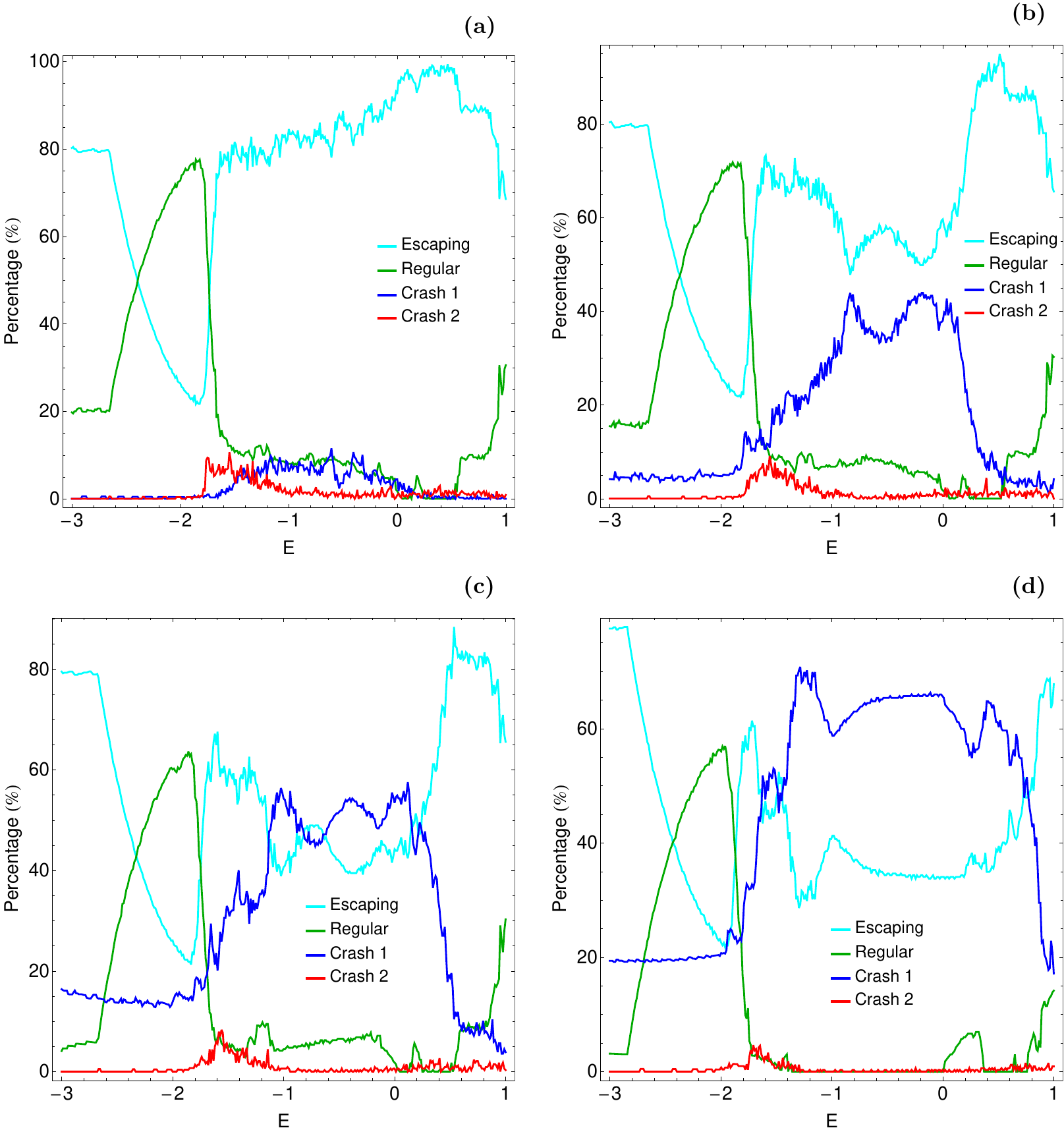}}
\caption{Evolution of the percentages of escaping, regular and collisional orbits on the $(x,E)$-plane as a function of the total orbital energy $E$. (a-upper left): $A_1 = 0$; (b-upper right): $A_1 = 0.001$; (c-lower left): $A_1 = 0.01$; (d-lower right): $A_1 = 0.1$.}
\label{pxE}
\end{figure*}

\begin{figure*}[!tH]
\centering
\resizebox{\hsize}{!}{\includegraphics{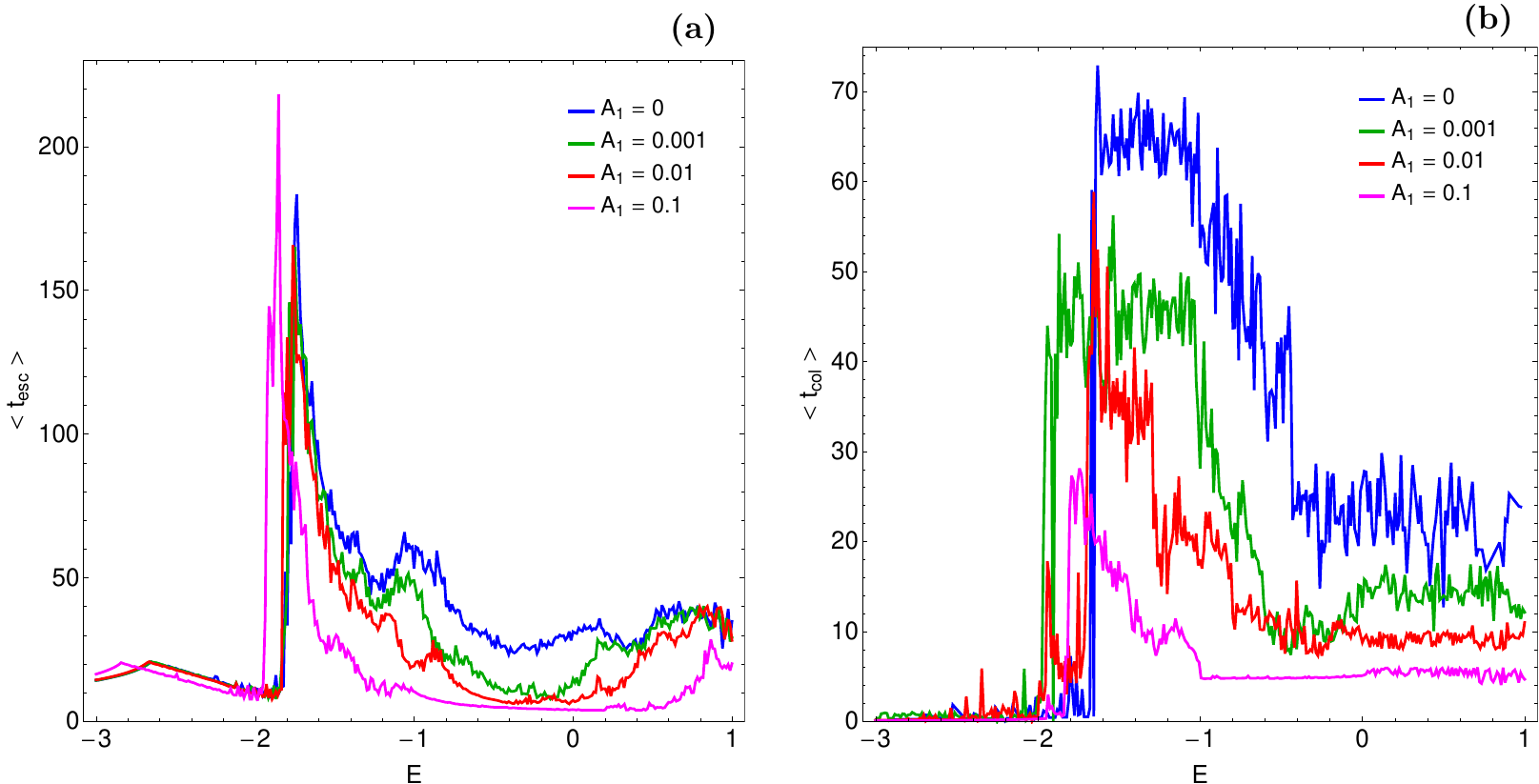}}
\caption{(a-left): The average escape time of orbits $< t_{\rm esc} >$ and (b-right): the average collision time of orbits $< t_{\rm col} >$ as a function of the total orbital energy $E$.}
\label{stats}
\end{figure*}

The color-coded OTDs in the configuration $(x,y)$ space provide sufficient information on the phase space mixing however, for only a fixed value of the energy integral and also for orbits that traverse the surface of section retrogradely. H\'{e}non \citep{H69}, introduced a new type of plane which can provide information not only about stability and chaotic regions but also about areas of bounded and unbounded motion using the section $y = \dot{x} = 0$, $\dot{y} > 0$ (see also \citet{BBS08}). In other words, all the initial conditions of the orbits of the test particles are launched from the $x$-axis with $x = x_0$, parallel to the $y$-axis $(y = 0)$. Consequently, in contrast to the previously discussed types of planes, only orbits with pericenters on the $x$-axis are included and therefore, the value of the energy $E$ can now be used as an ordinate. In this way, we can monitor how the energy influences the overall orbital structure of our dynamical system using a continuous spectrum of energy values rather than few discrete energy levels. In Fig. \ref{xE} we present the orbital structure of the $(x,E)$ plane for four values of the oblateness coefficient when $E \in [-3,1]$, while in Fig. \ref{xEt} the distribution of the corresponding escape and collision times of the orbits is depicted.

We can observe the presence of several types of regular orbits around the two primary bodies. Being more precise, on both sides of the primaries we identify stability islands corresponding to both direct (counterclockwise) and retrograde (clockwise) quasi-periodic orbits. It is seen that a large portion of the exterior region, that is for $x < x(L_3)$ and $x > x(L_2)$, a large portion of the $(x,E)$ plane is covered by initial conditions of escaping orbits however, at the left-hand side of the same plane two stability islands of regular orbits that circulate around both primaries are observed. Additional numerical calculations reveal that for much lower values of $x$ $(x < 5)$ these two stability islands are joined and form a crescent-like shape. We also see that collisional basins to oblate primary 1 leak outside the interior region, mainly outside $L_2$, and create complicated spiral shapes in the exterior region. On the other hand, the thin bands represent initial conditions of orbits that collide with primary body 2 are much more confined. It should be pointed out that in the blow-ups of the diagram several additional very small islands of stability have been identified\footnote{An infinite number of regions of (stable) quasi-periodic (or small scale chaotic) motion is expected from classical chaos theory.}.

As the value of the oblateness coefficient increases the structure of the $(x,E)$ planes exhibits the following changes: (i) The collisional basins to oblate primary 1 significantly grow and for $A_1 = 0.1$ they dominate; (ii) The area of the stability islands around primary 1 reduces and at the highest studied value of $A_1$ they completely disappear. According to Broucke's classification \citep{B68} the periodic orbits around the primaries belong to the families $C$ (at the left side of the primary) and $H_1$ (at the right side of the primary), while \citet{SS00} proved for the planar Hill's problem that the stability regions of the $C$ family are  more stable than those of the $H_1$ family. Our computations show that both families disappear for large values of the oblateness coefficient; (iii) The stability islands around primary body 2 are almost unperturbed, at least in the interval $E \in [-3, -1.5]$, by the shifting on the value of the oblateness. The phenomenon that stability islands can appear and disappear as a dynamical parameter is changed has also been reported in earlier paper (e.g., \citet{BBS06,dAT14}); (iv) Inspecting Figs. \ref{xEt}(a-d) we see that orbits with initial conditions very close to the center of the oblate primary 1 collide almost immediately with it, while their portion (or in other words the thickness of the vertical basin) increases for larger values of the oblateness coefficient $A_1$; (v) Another interesting phenomenon is the fact that as the primary body 1 becomes more and more oblate the fractality of the $(x,E)$ plane reduces and the boundaries between escaping and collisional basins appear to become smoother. It should be emphasized that the fractality of the structures was not measured by computing the corresponding fractal dimension. When we state that an area is fractal we mean that it has a fractal-like geometry.

It would be very informative to monitor the evolution of the percentages of all the different types of orbits as a function of the total orbital energy $E$ for the $(x,E)$ planes shown in Figs. \ref{xE}(a-d). Our results are presented in Figs. \ref{pxE}(a-d). We see that in all four cases the rate of the escaping orbits starts at about 80\% for $E = -3$ and gradually reduces until about 20\% for $E = -1.8$. For $E > -1.8$ it suddenly increases but this increase is controlled by the oblateness coefficient. In particular, the higher the value of $A_1$ the lower the increase of the escaping percentage in the energy interval $[-1.8, 1]$. In the same interval however, according to Fig. \ref{xEt}a escaping orbits dominate the $(x,E)$ plane when $A_1 = 0$. The evolution of the percentage of regular orbits displays similar patterns in the energy interval $[-3, -1.8]$ however, the corresponding rates are reduced with increasing $A_1$. For larger values of the energy the percentage of bounded orbits fluctuate and in general terms we may say that again the oblateness coefficient has an influence. The percentage of collisional orbits to oblate primary 1 has a monotone behaviour in the energy interval $[-3,-1.8]$, while for $E > -1.8$ it increases. This increase becomes stronger and stronger as the value of the oblateness coefficient increases. For large enough values of the energy the rate of collisional orbits to primary 1 decreases in all four cases, while at the same time escaping orbits is the most populated type of orbits. The percentage of collisional orbits to primary 2 on the other hand, is extremely low, while it peaks only in the energy interval $[-1.8, -1.2]$. Once more, the peaks in this energy range are influenced by the value of $A_1$; the larger the oblateness coefficient the lower the peaks.

Another interesting aspect would be to reveal how the oblateness coefficient influences the escape as well as the collision time of orbits. The evolution of the average value of the escape time $< t_{\rm esc} >$ of orbits as a function of the total orbital energy is given in Fig. \ref{stats}a. It is evident, especially in the energy interval $[-1.8, 1]$, that as the value of the oblateness coefficient increases the escape rates of orbits are reduced. In the same vein in Fig. \ref{stats}b we present the evolution of the average collision time $< t_{\rm col} >$ of orbits that collide to oblate primary 1. Once more we observe that the collision time of orbits decreases with increasing value of $A_1$, especially in the energy interval $[-1.6, 1]$. Additional numerical calculations, not shown here, indicate that the oblateness coefficient also influences the collision rates of orbits which collide to spherical primary 2. In particular, there are energy ranges in which the collision rates to primary 2 are reduced with increasing $A_1$, however the diagrams are not so clear as in Fig. \ref{stats}b, so we decided not to include them.

Before closing this section we would like to add that the particular value of the mass ratio $\mu$ does not really change the qualitative nature of the numerical outcomes presented in this section. Indeed after conducting some additional calculations with larger and lower values of $\mu$ we concluded to the same results. The parameters which mostly influence the orbital dynamics are the total energy and of course the oblateness coefficient.

\section{Conclusions}
\label{conc}

The aim of this work was to reveal how the oblateness coefficient influences the character of orbits in the classical planar circular restricted three-body problem. After conducting an extensive and thorough numerical investigation we managed to distinguish between bounded, escaping and collisional orbits and we also located the basins of escape and collision, finding also correlations with the corresponding escape and collision times. Our numerical results strongly suggest that the oblateness coefficient plays a very important role in the nature of the test's body motion under the gravitational field of two primaries. To our knowledge, this is the first detailed and systematic numerical analysis on the influence of the oblateness coefficient on the character of orbits and this is exactly the novelty and the contribution of the current work.

For several values of the oblateness coefficient in the last three Hill's regions configurations we defined dense uniform grids of $1024 \times 1024$ initial conditions regularly distributed on the $\dot{\phi} < 0$ part of the configuration $(x,y)$ plane inside the area allowed by the value of the total orbital energy. All orbits were launched with initial conditions inside the scattering region, which in our case was a square grid with $-2\leq x,y \leq 2$. For the numerical integration of the orbits in each type of grid, we needed about between 11 hours and 5 days of CPU time on a Pentium Dual-Core 2.2 GHz PC, depending on the escape and collisional rates of orbits in each case. For each initial condition, the maximum time of the numerical integration was set to be equal to $10^4$ time units however, when a particle escaped or collided with one of the two primaries the numerical integration was effectively ended and proceeded to the next available initial condition.

In this article we provide quantitative information regarding the escape and collisional dynamics in the restricted three-body problem with oblateness. The main numerical results of our research can be summarized as follows:
\begin{enumerate}
 \item It was observed that as the value of the oblateness coefficient increases the size of the stability islands around primary 1 is reduced and at relatively high values of the oblateness coefficient there is no indication of regular motion around the oblate primary, while at the same time the stability region around spherical primary 2 is also reduced.
 \item The collisional basins to primary 1 were found to significantly increase in size with increasing value of the oblateness coefficient, while the basins formed by initial conditions of orbits which collide to the primary 2 are practically unaffected by the shifting on the value of the oblateness coefficient.
 \item It was detected that in the vicinity of the center of the oblate primary 1 a portion of orbits collide to the same primary almost immediately. Furthermore, the amount of these type of orbits increases with increasing oblateness. On the other hand this behaviour do not apply to the case of the spherical primary 2.
 \item Our calculations reveal that as the primary body 1 becomes more and more oblate the fractality of the planes is reduced and the boundaries between the escaping and collisional basins appear to become smoother.
 \item We presented evidence that the oblateness coefficient also influences the escape as well as the collision time of the orbits. In particular, both types of times are reduced as the value of the oblateness coefficient increases.
\end{enumerate}

Judging by the detailed and novel outcomes we may say that our task has been successfully completed. We hope that the present numerical analysis and the corresponding results to be useful in the field of escape dynamics in the restricted three-body problem with oblateness. The outcomes as well as the conclusions of the present research are considered, as an initial effort and also as a promising step in the task of understanding the escape mechanism of orbits in this interesting version of the classical three-body problem. Taking into account that our results are encouraging, it is in our future plans to properly modify our dynamical model in order to expand our investigation into three dimensions and explore the entire six-dimensional phase thus revealing the influence of the oblateness coefficient on the orbital structure.

\section*{Acknowledgments}

I would like to express my warmest thanks to the anonymous referee for the careful reading of the manuscript and for all the apt suggestions and comments which allowed us to improve both the quality and the clarity of the paper.

\end{document}